\documentclass[notitlepage,nofootinbib,aps,prd,showpacs,onecolumn,
superscriptaddress,article]{revtex4-2}

\usepackage{graphicx}
\usepackage{dcolumn}
\usepackage{amsmath}
\usepackage[makeroom]{cancel}
\usepackage{amssymb}
\usepackage{bm}
\usepackage{color}
\usepackage[normalem]{ulem}
\usepackage[dvipsnames]{xcolor}
\usepackage{hyperref}
\usepackage[many]{tcolorbox}

\hypersetup{
colorlinks=true, 
linkcolor=blue,  
citecolor=cyan,  
}
\usepackage{tikz}
\makeatletter
\newcommand*{\encircled}[1]{\relax\ifmmode\mathpalette\@encircled@math{#1}\else\@encircled{#1}\fi}
\newcommand*{\@encircled@math}[2]{\@encircled{$\m@th#1#2$}}
\newcommand*{\@encircled}[1]{%
  \tikz[baseline,anchor=base]{\node[draw,circle,outer sep=0pt,inner sep=.2ex] {#1};}}
\makeatother

\usetikzlibrary{arrows}
\usetikzlibrary{shapes}

\newtcolorbox{cross}{blank,breakable,parbox=false,
  overlay={\draw[blue,line width=2pt] (interior.south west)--(interior.north east);
    \draw[blue,line width=2pt] (interior.north west)--(interior.south east);}}

\begin{document}
\title{Tidal heating in a Riemann--Cartan spacetime}

\author{Sudipta Hensh}
\email{f170656@fpf.slu.cz, sudiptahensh2009@gmail.com}
\affiliation{Research Centre for Theoretical Physics and Astrophysics, Institute of Physics, Silesian University in Opava, Bezru\v{c}ovo n\'{a}m\v{e}st\'{i} 13, CZ-74601 Opava, Czech Republic}

\author{Stefano Liberati}
\email{liberati@sissa.it}
\affiliation{SISSA - International School for Advanced Studies, Via Bonomea 265, 34136 Trieste, Italy }
\affiliation{IFPU - Institute for Fundamental Physics of the Universe, Via Beirut 2, 34014 Trieste, Italy }
\affiliation{INFN Sezione di Trieste, Via Valerio 2, 34127 Trieste, Italy }

\author{Vincenzo Vitagliano}
\email{vincenzo.vitagliano@unige.it}
\affiliation{DIME, University of Genova, Via all’Opera Pia 15, 16145 Genova, Italy}
\affiliation{INFN Sezione di Genova, Via Dodecaneso 33, 16146 Genova, Italy,}

\date{\today}

\begin{abstract}
We consider possible perturbations of the black hole event horizon induced by matter with spin, extending the derivation of the Hawking--Hartle formula  (tidal heating) in the presence of torsion. When specialised to theories with a non-vanishing (pseudo-)traceless component of the (con)torsion tensor, we remarkably find that the tidal heating phenomenon gets modified by additional torsion-dependent terms, in agreement with previous investigations based on Jacobson's spacetime thermodynamics approach. These results lead to relevant phenomenological and theoretical consequences: modifications in the Hawking--Hartle term change the Bondi mass associated with the gravitational radiation observed at infinity, and modify the Hawking radiation spectrum of evaporating black holes.

\end{abstract}


\maketitle

\section{Introduction}

In statistical physics, one of the consequences of Onsager's principle of reciprocity when it comes to irreversible processes is the regression hypothesis~\cite{PhysRev.37.405, PhysRev.38.2265}: the evolution of spontaneous fluctuations of a dissipative system is governed by the macroscopic equations that determine how the same system approaches its equilibrium state after being externally perturbed (in a regime of linear response). This result is universal: while the selection of forces and fluxes to which the Onsager relations apply can be tricky, it is nevertheless surprisingly indifferent to the details of the thermodynamic system once the dynamical assumptions hold. Onsager's regression hypothesis is a classical hypothesis, the quantum-mechanical generalising counterpart being the fluctuation-dissipation theorem, initially proven by Callen and Welton~\cite{PhysRev.83.34}, and later refined by Kubo~\cite{Kubo:1957mj}.

In a seminal paper by Candelas and Sciama~\cite{PhysRevLett.38.1372}, the fluctuation-dissipation theorem has been proven to apply also to the context of black holes irreversible and non-equilibrium thermodynamics, according to the following picture. The effect of the coupling between a rotating black hole and a companion compact object becomes manifest through the {\em tidal heating} phenomenon~\cite{Hartle:1973zz,Hawking:1972hy,Hughes:2001jr}: during the in-spiral phase of the binary evolution, the two bodies deform each other through their respective tidal fields, making the mass and angular momentum of each member evolving with time. Tidal heating has been the subject of intense studies up to recent years~\cite{Hughes:2001jr,Alvi:2001mx,Chatziioannou:2012gq,Isoyama:2017tbp,Datta:2019epe} as it can unveil signatures associated with alternative theories of gravity as well as with general relativity e.g. in the presence of black holes mimickers (for example, the tidal heating for exotic compact objects was studied in~\cite{Datta:2019epe,Datta:2019euh,Maselli:2017cmm,Datta:2020gem,Datta:2020rvo,Datta:2021row,Chakraborty:2021gdf}). In standard (Riemannian) general relativity, a textbook calculation (see for example~\cite{Chandrasekhar:1985kt}) relates the rate of change of the horizon area $\Sigma$ of the rotating black hole to the changes in mass and angular momentum, and subsequently to the first-order perturbation in the shear $\sigma$ at the horizon; the result is the well known {\em Hawking--Hartle area formula}, which, given the proportionality between area and black hole entropy, provides also the rate of dissipation of the gravitational perturbation: a rotating black hole with $\left.\sigma\right|_{\text{hor}}\neq0$, radiating gravitational waves, reduces the shear and approaches a stationary state. However, according to the fluctuations-dissipation theorem, we could promote the shear $\sigma$ to the role of quantum operator, and describe the relaxation of the black hole to the stationary configuration as if it were due to the Hawking phenomenon for the shear fluctuations. 

Embracing Candelas and Sciama's reasoning, one can then infer that any fundamental mechanism with the capability of altering the way the black hole dissipates the gravitational perturbations - {\em viz.} any mechanism inducing a modification in the Hawking--Hartle formula - will also in turn tweaks the corresponding black hole Hawking emission spectrum. In the present paper, we start our program in this direction.

General relativity considers {\em a priori} the affine connection to be symmetric and metric compatible (Levi-Civita connection); we will focus on a quite conservative extension of this {\em ansatz}, starting from the realisation that the symmetry condition can be relaxed, allowing for the connection to contain also an anti-symmetric part, the so-called torsion tensor. In this work, we will calculate for the first time, to the best of our knowledge, how tidal heating gets modified by contributions of torsional origin in a $U_4$ (Riemann--Cartan) spacetime, following a completely generic approach. 
In terms of geometry, the prototypical and most natural example of modification of general relativity allowing for non-vanishing torsion is the Einstein--Cartan(--Sciama--Kibble) theory. Despite its simplicity, there is reason to expect the Hawking--Hartle tidal heating formula to be modified even in this setting: in a recent derivation of Einstein--Cartan field equations~\cite{Dey:2017fld} (see also \cite{Dey:2022fll,Dey:2022qqp}) from Jacobson's spacetime thermodynamics approach~\cite{Jacobson:1995ab}, the analysis of the irreversible part of the Clausius equation showed that the generalised Hawking--Hartle formula should acquire an additional term, proportional to the twist of the horizon congruence; this is due to the fact that, when torsion is present, hypersurface orthogonality of the congruence does not imply a vanishing twist. Here, we shall explicitly verify this circumstantial evidence by a direct calculation and discuss its theoretical and observational implications.

\section{Newman--Penrose Formalism and Optical Scalars: the Torsionfull Case \label{sec2}}
The standard calculation of the tidal heating formula relies on the equation governing the evolution of the expansion - the real part of the complex divergence, {\em aka} first Sachs' optical parameter - of a null geodesics congruence. While geodesic deviation and the associated Raychaudhuri equation in a spacetime with torsion have been extensively and consistently studied in the literature (recent developments can be found in \cite{Luz:2017ldh,Iosifidis:2018diy,Cembranos:2018ipn,Hensh:2021oyk}), the generalisation to the full set of optical equations, initially obtained more than forty years ago in~\cite{Jogia1980,Jogia81} with surprisingly limited success, has been re-discovered only quite recently~\cite{Speziale:2018cvy}. In order to incorporate torsion effects, it is necessary, in the first instance, to generalise the Newman--Penrose spin coefficients formalism\footnote{Troughout the paper, we consider $G=c=1$, signature $(+,-,-,-)$, and indices running from $1$ to $4$. Note also that we follow the conventions of \cite{Jogia1980,Jogia81}, which is slightly different from the traditional textbook one \cite{Chandrasekhar:1985kt} for what concerns the Ricci rotation coefficients, but that agrees with the definition of the optical scalars.}. Let $\{l^{\mu}, n^{\mu}, m^{\mu}, m^{*\mu}\}$ be a null tetrad consisting of a pair of real vectors ($l^{\mu}$ and $n^{\mu}$) and a pair of complex-conjugate vectors ($m^{\mu}$ and $m^{*\mu}$). The two pairs are usually constrained to satisfy normalisation conditions, $l^{\mu} n_{\mu} = 1  $ and $ m^{\mu} m^*_{\mu} = -1$\,. For the sake of conciseness, we introduce the notation 
\begin{eqnarray}\label{tet}
    e_{(1)}^{\,\,\,\,\,\,\mu} \equiv l^{\mu} \ , \quad e_{(2)}^{\,\,\,\,\,\,\mu} \equiv n^{\mu} \ , \quad e_{(3)}^{\,\,\,\,\,\,\mu} \equiv m^{\mu} \ , \quad 
    e_{(4)}^{\,\,\,\,\,
    \,\mu} \equiv m^{*\mu} \,.
\end{eqnarray}
In the following, numbers enclosed in parentheses will refer to the corresponding tetrad vector in \eqref{tet}; moreover, we shall generally use Latin letters for the tetrad components and Greek ones for the tensorial ones. Using the normalisation conditions and the fact that the norms of the four vectors vanish, the constant symmetric matrix $\eta_{a b} \equiv e_{a}^{\,\,\,\mu} e_{b \mu}$ - the object lowering and raising tetrad indices - takes the form
\begin{eqnarray}
    \eta_{a b} = \eta^{a b} = \begin{pmatrix} 0 & 1 & 0 & 0\\ 1 & 0 & 0 & 0 \\ 0 & 0 & 0 & -1 \\ 0 & 0 & -1 & 0 \end{pmatrix} \ .
\end{eqnarray}

In order to use the Newman--Penrose formalism, the approach is to express any physical quantity in terms of tetrad components. According to the fact that covariant derivatives are now encoding a contribution coming from torsion, one can show that the Ricci rotation coefficients read $    \widehat{\gamma}_{abc} \equiv e_c^{\,\,\,\nu} e_b^{\,\,\,\mu} \widehat{\nabla}_{\nu} e_{a\mu}
    = \gamma_{abc} + K_{cba} $\,,
where, from now on, hatted quantities are calculated with respect to the generic (torsionfull) connection, whereas terms without hat refer to the Christoffel part. The tensor $K_{abc}$ on the right-hand side is the {\em contorsion} tensor (in our notation $T_{abc}$ stays instead for the torsion tensor, see also the Appendix). In order to deal with this generalised approach to the Newman--Penrose formalism, we define two sets of scalars, encompassing respectively the sole Riemannian (those without the subscript T, which are equivalent to the usual optical scalar of general relativity) and torsion (with the subscript T) effects,
\begin{equation}\label{opt_scalars}
\def\arraystretch{1.3}
\begin{array}{@{}lll@{}}
    &  \kappa = l^{\nu} m^{\mu} \nabla_{\nu} l_{\mu} \, , 
     &  \kappa_{_T} = K_{\lambda \mu \nu} l^{\lambda} m^{\mu} l^{\nu} \,,\\
     & \rho = m^{*\nu} m^{\mu} \nabla_{\nu} l_{\mu} \, ,  
      &  \rho_{_T} = K_{\lambda \mu \nu} m^{*\lambda} m^\mu l^\nu  \, ,\\
          & \sigma = m^{\nu} m^{\mu} \nabla_{\nu} l_{\mu} \, ,     
     &   \sigma_{_T}= K_{\lambda \mu \nu} m^{\lambda} m^{\mu} l^{\nu} \, ,\\
          &  \tau = n^{\nu} m^{\mu} \nabla_{\nu} l_{\mu} \, , 
     &  \tau_{_T} = K_{\lambda \mu \nu} n^{\lambda} m^{\mu} l^{\nu} \,,\\
          &  \pi = -l^{\nu} m^{*\mu} \nabla_{\nu} n_{\mu} \, , 
     &  \pi_{_T} = -K_{\lambda \mu \nu} l^{\lambda} m^{*\mu} n^{\nu} \,,\\
          & \epsilon = \frac{1}{2} l^{\nu} \left( n^{\mu} \nabla_{\nu} l_{\mu} - m^{*\mu} \nabla_{\nu} m_{\mu} \right) \, , \qquad\qquad
      &   \epsilon_{_T}= \frac{1}{2} K_{\lambda \mu \nu} l^{\lambda} \left( n^\mu l^\nu - m^{*\mu} m^{\nu} \right) \, ,\\
                & \alpha = \frac{1}{2} m^{*\nu} \left( n^{\mu} \nabla_{\nu} l_{\mu} - m^{*\mu} \nabla_{\nu} m_{\mu} \right) \, , \qquad\qquad
      &   \alpha_{_T}= \frac{1}{2} K_{\lambda \mu \nu} m^{*\lambda} \left( n^\mu l^\nu - m^{*\mu} m^{\nu} \right) \, ,\\
                & \beta = \frac{1}{2} m^{\nu} \left( n^{\mu} \nabla_{\nu} l_{\mu} - m^{*\mu} \nabla_{\nu} m_{\mu} \right) \, , \qquad\qquad
      &   \beta_{_T}= \frac{1}{2} K_{\lambda \mu \nu} m^{\lambda} \left( n^\mu l^\nu - m^{*\mu} m^{\nu} \right) \, .
\end{array}
\end{equation}

We will keep whenever possible the same notation of the standard Riemannian case: then, for example, the quantity $\rho$ is split into two contributions, $\rho=-\theta+ \imath~ \omega$, the first being the expansion $\theta$ (corresponding to minus the real part of $\rho$) and the second being the twist $\omega$ (the imaginary part of $\rho$). 
Keeping in mind the definitions \eqref{opt_scalars}, it is possible to analyse the properties of geodesy and hypersurface orthogonality of a congruence of $l^\mu$ vectors. In general relativity, the analysis of the propagation of tetrad vectors along $l^\mu$ proves that its direction defines a congruence of affinely parametrised null geodesics iff $\kappa=\Re(\epsilon)=0$; in addition, with a suitable rotation, the remaining basis vectors parallelly propagate along $l^\mu$. In the torsionfull case, the first-order change in a basis vector $e_{a}$ under an infinitesimal displacement $\xi$ reads 
\begin{equation} \label{eq:21}
\delta e_{a \mu} = \xi^\nu\widehat{\nabla}_\nu e_{a \mu}  = e^{b}_{\,\,\,\mu} \widehat{\gamma}_{a b c} e^{c}_{\,\,\,\nu} \xi^\nu = \widehat{\gamma}_{a b c} e^{b}_{\,\,\,\mu} \xi^{c} \, .
\end{equation}
Using~\eqref{eq:21} together with~\eqref{opt_scalars} and the definition of the rotation coefficients, we can write $D l_\mu$  -- the change in $l_\mu$ per unit displacement along the direction of the first basis vector, i.e. $l_\mu$ itself -- as
\begin{eqnarray} \label{eq:22bis}
D l_\mu 
= (\epsilon + \epsilon^*) l_\mu + (\epsilon_{_T} + \epsilon^*_{_T}) l_\mu - (\kappa+\kappa_{_T}) m^*_\mu - (\kappa^* + \kappa^*_{_T})m_\mu  \,.
\end{eqnarray}
This expression provides the requirements for the $l_\mu$ vectors to form a congruence of geodesics, which happens when {\em both} the conditions $\kappa=l^{\nu} m^{\mu} \nabla_{\nu} l_{\mu}=0$, {and} $\kappa_{_T}=K_{\lambda \mu \nu} l^{\lambda} m^{\mu} l^{\nu} = 0$ are fulfilled. The condition $\kappa+\kappa_{_T}=0$ defines a congruence of autoparallel curves. Furthermore, if geodesics are affinely parametrized then it should also follow $\epsilon + \epsilon^*=2~\Re(\epsilon) = 0$ and $\epsilon_{_T} + \epsilon^*_{_T}=2~\Re(\epsilon_{_T}) = 0$.
The last remark concerns the congruence property of hypersurface orthogonality. In the presence of torsion, a hypersurface orthogonal vector field with tangent vector given by $\chi^{\alpha}$ satisfies
\begin{eqnarray}\label{hyper}
\chi_{[\lambda}\widehat{\nabla}_{\nu} \chi_{\mu]} = \frac{1}{3!} \left(-2 \chi_{\lambda} T_{\nu \mu}^{\,\,\,\,\,\,\gamma} \chi_{\gamma} -2 \chi_{\nu} T_{\mu \lambda}^{\,\,\,\,\,\,\gamma} \chi_{\gamma} -2 \chi_{\mu} T_{\lambda \nu}^{\,\,\,\,\,\,\gamma} \chi_{\gamma}\right) \,,
\end{eqnarray}
where the right-hand side clearly vanishes when there is no torsion (for a useful discussion see \cite{Luz:2019kmm}).  On the other hand, from the torsionfull covariant derivative $   \widehat{\nabla}_\nu l_{\mu} = e^{b}_{\,\,\,\mu} \widehat{\gamma}_{(1) b c} e^{c}_{\,\,\,\nu} $, we find~\footnote{Here we are assuming $l^\mu$ to define a congruence of autoparallel curves, viz $\kappa+\kappa_{_T}=0$ again; otherwise, \eqref{lll} would contain the extra terms $-(\kappa^* m_{[\mu} n_\nu+\kappa m^*_{[\mu} n_\nu) l_{\lambda]}-(\kappa_{_T}^* m_{[\mu} n_\nu+\kappa_{_T} m^*_{[\mu} n_\nu) l_{\lambda]}$}
\begin{eqnarray}\label{lll}
l_{[\lambda}\widehat{\nabla}_\nu l_{\mu]} = (\rho - \rho^*) m^*_{[\mu}m_{\nu} l_{ \lambda]} + (\rho_{_T} - \rho^*_{_T}) m^*_{[\mu}m_{\nu} l_{\lambda]}\, .
\end{eqnarray}
If torsion is absent, the second term on the right-hand side of \eqref{lll} is trivially zero, since, using \eqref{opt_scalars}, $\rho_{_T} = \rho^*_{_T}=0$; in the Riemannian case, then, the condition for $l^{\alpha}$ to be tangent to a hypersurface orthogonal congruence implies
\begin{eqnarray}
l_{[\lambda}\nabla_{\nu}l_{\mu]} =0= (\rho - \rho^*) m^*_{[\mu}m_{\nu} l_{ \lambda]}  \, , \label{eq:27}
\end{eqnarray}
which in turn is true iff $\rho=\rho^*$, that is, comparing again with \eqref{opt_scalars}, if the congruence is irrotational (no twist). However, when torsion is switched on, the term proportional to $\rho_{_T} - \rho^*_{_T}=2 T_{\mu \nu}^{\,\,\,\,\,\,\gamma} m^{*\mu} m^{\nu} l_\gamma$, in general, does not vanish. In this case, the hypersurface orthogonality of a congruence of autoparallel curves does not guarantee a vanishing twist.

%
The equations governing the evolution of optical parameters may be inferred by applying the Ricci identity (see \eqref{RicciId} in the Appendix) to the four tetrad vectors and taking the tetrad components of the resulting expression. We obtain
\begin{eqnarray}\label{tor_hyp2}
  \widehat{\nabla}_d \widehat{\gamma}_{abc} - \widehat{\nabla}_c \widehat{\gamma}_{abd}&=& \widehat{\gamma}_{qac}\widehat{\gamma}^{q}{}_{bd}-\widehat{\gamma}_{qad}\widehat{\gamma}^{q}{}_{bc}-2 \widehat{\gamma}_{ab}{}^q(\widehat{\gamma}_{q[cd]}-K_{[dc]q})+ \widehat{R}_{cdab} \,.
\end{eqnarray}
Here $\widehat{R}_{abcd}=\widehat{R}_{\mu\nu\rho\sigma}e_{a}{}^{\mu}e_{b}{}^{\nu}e_{c}{}^{\rho}e_{d}{}^{\sigma}$  are the components of the Riemann tensor of the full connection in the tetrad basis. Different choices of the four tetrad indices $\{abcd\}$ will correspond to different evolution equations for the various optical scalars. Since we are interested in the equation for $\widehat{\rho}$, we specialise the previous identity to the combination $\{abcd\}\rightarrow \{(1)(3)(4)(1)\}$. In this case, in fact, being $D$ and $\delta^*$ the derivatives along, respectively, $l^\mu$ and $m^{*\mu}$, we get
\begin{eqnarray} \label{eq:B1}
    D \widehat{\rho} - \delta^* \widehat{\kappa} & = & \widehat{\rho}~ ( \widehat{\rho} + \widehat{\epsilon} + \widehat{\epsilon}^* ) + \widehat{\sigma} \widehat{\sigma}^* - \widehat{\tau} \widehat{\kappa}^* - \widehat{\kappa} (3 \widehat{\alpha} + \widehat{\beta}^* - \widehat{\pi} ) 
    - \widehat{\rho} (\rho_{_T} - \epsilon_{_T} + \epsilon^*_{_T} ) - \widehat{\sigma} \sigma^*_{_T} + \widehat{\tau} \kappa^*_{_T} \nonumber \\
    && + \widehat{\kappa} ~(3~ \alpha_{_T} + \beta^*_{_T} - \pi_{_T} ) + \widehat{R}_{(4)(1)(1)(3)} \,.
\end{eqnarray}
In our notation, the hatted optical scalars are the sum of the two homonymous Riemannian and torsion contributions from \eqref{opt_scalars} (for example, $\widehat{\rho}=\rho+\rho_{_T}$). 
Written in terms of the generalised optical scalars set \eqref{opt_scalars} - once the Riemann tensor is disentangled into its metric and contorsion parts - the expression \eqref{eq:B1} reads
\begin{eqnarray}\label{12new}
D\rho=\rho^2+2~\rho~\Re(\epsilon)+\left|\sigma\right|^2+\Phi_{00} 
+\delta^* \kappa  
+\kappa(\pi-3\alpha-\beta^*)-\tau\kappa^*\,,
\end{eqnarray}
where $ \Phi_{00}= - \frac{1}{2} R_{\mu \nu} l^{\mu} l^{\nu}$. We are interested in a congruence of null autoparallel trajectories - not necessarily affinely parametrised - generated by $l^\mu$, $l^\nu\widehat{\nabla}_\nu l^\mu=k l^\mu$. In this case, $\widehat{\kappa}=0\rightarrow \kappa=-\kappa_{_T}$. We can then recast \eqref{12new} as
\begin{eqnarray}\label{13new}
D\rho=\rho^2+2~\rho~\Re(\epsilon)+\left|\sigma\right|^2+\Phi_{00} 
-\delta^* \kappa_T
+\kappa_T(3\alpha+\beta^*-\pi)+\tau\kappa_{_T}^*\,,
\end{eqnarray}
which is also consistent with previous results \cite{Jogia1980,Jogia81,test1}. Note that for $\kappa=0$ (that is, when autoparallel curves are also metric geodesics), one gets consequently $\kappa_{_T}=0$ and recovers the usual $V_4$ (Riemannian space-time of general relativity) result
\begin{eqnarray}
D\rho=\rho^2+2~\rho~\Re(\epsilon)+\left|\sigma\right|^2+\Phi_{00}\,. 
\end{eqnarray}
A relevant question one might ask is whether all the components of the torsion tensor democratically contribute to the modified expression \eqref{13new}; as the reader should expect, the answer is negative. In order to prove this statement, it is customary to introduce the decomposition of contorsion into its three irreducible parts with respect to the Lorentz group,
\begin{eqnarray}\label{irrep}
K^{\mu\nu\rho}={}^{(1)}\!K^{\mu\nu\rho}+\frac{2}{3}g^{\mu[\rho}~ {}^{(2)}\!K^{\nu]}+\varepsilon^{\mu\nu\rho\sigma} {}^{(3)}\!K_{\sigma}\,, 
\end{eqnarray}
where $\varepsilon_{\mu\nu\rho\sigma}=-4!~\imath~ l_{[\mu} n_{\nu} m_{\rho} m_{\sigma]}^*$ is the Levi-Civita tensor, and we have defined a trace vector ${}^{(2)}\!K^{\nu}=K_{\mu}{}^{\nu\mu}$ (corresponding to 4-dim
irreducible representation $\left(\frac{1}{2},\frac{1}{2}\right)$), an axial vector ${}^{(3)}\!K^{\sigma}=\frac{1}{6}\varepsilon^{\sigma\mu\nu\rho}K_{\mu\nu\rho}$ (also $\left(\frac{1}{2},\frac{1}{2}\right)$), and a (pseudo-)traceless tensor ${}^{(1)}\!K^{\mu\nu\rho}$ which satisfies $g_{\mu\nu}{}^{(1)}\!K^{\mu\nu\rho}=0=\varepsilon_{\sigma\mu\nu\rho}{}^{(1)}\!K^{\mu\nu\rho}$ (corresponding to the 16-dim irreducible representation $\left(\frac{3}{2},\frac{1}{2}\right)\oplus\left(\frac{1}{2},\frac{3}{2}\right)$). Using the properties of the tetrad vectors, combined with \eqref{irrep}, we get 
\begin{eqnarray}\label{kt}
 \kappa_{_T} = K^{\mu\nu\rho} l_{\mu} m_{\nu} l_{\rho} =\left({}^{(1)}\!K^{\mu\nu\rho}+\frac{2}{3}g^{\mu[\rho}~ {}^{(2)}\!K^{\nu]}+\varepsilon^{\mu\nu\rho\sigma} {}^{(3)}\!K_{\sigma}\right) l_{\mu} m_{\nu} l_{\rho}={}^{(1)}\!K^{\mu\nu\rho}~l_{\mu} m_{\nu} l_{\rho}\,.
\end{eqnarray}
According to \eqref{kt}, and in order to have a non-trivial effect on the evolution of $\rho$, see \eqref{13new}, (con)torsion has to have a non-zero ${}^{(1)}\!K^{\mu\nu\rho}$ component (which usually is referred to in the literature as the `spin-2' component). This is a rather crucial condition, ruling out several of the simplest models on the market that contain only the vector and/or axial components in \eqref{irrep}. The striking example is given by Einstein--Cartan theory with a minimally coupled Dirac field; in this case, the contorsion (and torsion as well) is a completely anti-symmetric tensor~\cite{Hehl:1976kj}, meaning that \eqref{irrep} collapses to the sole axial contribution and leading to $\kappa_T^{EC}=0$. Noteworthy, one could think to excite the ${}^{(1)}\!K^{\mu\nu\rho}$ component even in the context of Einstein--Cartan, for example by introducing non-minimal couplings (following the philosophy of \cite{Magueijo:2012ug, Benedetti:2011nd}, although the specific non-minimal coupling taken into account there would source only the trace part $ {}^{(2)}\!K^{\mu}$ and consequently would still result in a vanishing $\kappa_T$) or, more in general, in the context of the Standard Model Extension.

\section{Tidal heating with torsion \label{sec4}}

Our goal is to calculate the rate of change of the horizon area in the Hawking--Hartle basis \cite{Chandrasekhar:1985kt}, assuming for simplicity a Kerr black hole as the underlying solution. This can be surely done in the case of theories for which torsion does not propagate extra degrees of freedom \cite{Vitagliano:2010pq,Vitagliano:2010sr,Vitagliano:2013rna} since Kerr black hole will be still a vacuum solution. For theories with dynamical torsion, the following discussion will be valid modulo the extra assumption that the chosen solution does not step away too far from Kerr (subleading contributions from torsion to the solution, and more specifically to $\epsilon$). 

Splitting \eqref{13new} (and all the Newman--Penrose scalar quantities therein) into its real and imaginary part, one gets a first equation for the evolution of the expansion and a second one for the twist, both including now extra terms due to the torsional degrees of freedom,
\begin{eqnarray}
    D \theta 
    &=& - \theta^2 
    + 2 \theta \epsilon^{\mathrm{R}}
    + \omega^2
    - |\sigma|^2
    + \Phi_{00} 
    + A_{_T} \ , \label{eq18} \\
    D \omega 
    &=& -2 \theta \omega 
    + 2 \omega \epsilon^{\mathrm{R}}
    + B_{_T} \ \label{eq19} \ ,
\end{eqnarray}
where  $D=d/dv$, $v$ is the time parameter parametrising the trajectories, and
\begin{eqnarray}
    A_{_T} &=&     - 3 \alpha^{\mathrm{R}} \kappa_{_T}^{\mathrm{R}}
    - \beta^{\mathrm{R}} \kappa_{_T}^{\mathrm{R}} 
    + \pi^{\mathrm{R}} \kappa_{_T}^{\mathrm{R}} 
    + 3 \alpha^{\mathrm{I}} \kappa_{_T}^{\mathrm{I}}
    - \beta^{\mathrm{I}} \kappa_{_T}^{\mathrm{I}}
    - \pi^{\mathrm{I}} \kappa_{_T}^{\mathrm{I}} 
     - \tau^{\mathrm{R}} \kappa_{_T}^{\mathrm{R}}
    - \tau^{\mathrm{I}} \kappa_{_T}^{\mathrm{I}}
    +\delta^* \kappa_{_T}^{\mathrm{R}}\ , \\
    B_{_T} &=& 3 \alpha^{\mathrm{I}} \kappa_{_T}^{\mathrm{R}}
    - \beta^{\mathrm{I}} \kappa_{_T}^{\mathrm{R}} 
    - \pi^{\mathrm{I}} \kappa_{_T}^{\mathrm{R}}
    + 3 \alpha^{\mathrm{R}} \kappa_{_T}^{\mathrm{I}}
    +\beta^{\mathrm{R}} \kappa_{_T}^{\mathrm{I}}
    - \pi^{\mathrm{R}} \kappa_{_T}^{\mathrm{I}}  
    + \tau^{\mathrm{I}} \kappa_{_T}^{\mathrm{R}}
    - \tau^{\mathrm{R}} \kappa_{_T}^{\mathrm{I}}
    - \delta^* \kappa_{_T}^{\mathrm{I}} \ .
\end{eqnarray}

We assume that an influx of torsion excites the horizon as a pure gravitational perturbation. A typical physical scenario culminating in this situation can be pictured as follows: in a black hole -- compact object binary, the constituents of the inspiralling companion can be realistically thought of as particles endowed with spin, hence generating torsion; when the compact object falls onto the black hole, the torsion field intrinsically entangled within matter will further affect the horizon; this is the case, for example, of ultra-dense neutron stars plunging radially into a black hole. 

The first step to finding a mathematical description of the previous picture is to describe the Kerr solution in the Newman--Penrose formalism. For this purpose, it is necessary to write the metric in terms of an appropriate null tetrad (satisfying the conditions outlined at the beginning of Sec.~\ref{sec2}) and calculate from there the optical parameters. Kinnersley found a convenient tetrad \cite{Kinnersley:1969zza} by imposing $\kappa= \sigma= \lambda= \nu=0$ and using the remaining gauge freedom to set $\epsilon=0$. In general, one can pass from one tetrad basis to another by using appropriate transformations. We will make use of what Chandrasekhar defines as a type III rotation, namely a transformation $l^\mu\rightarrow \mathcal{A}^{-1} \cdot l^\mu, n^\mu\rightarrow \mathcal{A} \cdot n^\mu, m^\mu\rightarrow e^{\imath~\vartheta}\cdot m^\mu$, where $\mathcal{A}$ and $\vartheta$ are scalar functions. The Hawking--Hartle basis can be then deduced from the Kinnersley tetrad by performing a type III rotation with $\mathcal{A}=2 (r^2+a^2)/ (r^2-2 M r+a^2)$ and $\vartheta=0$. The Newman--Penrose scalars in the Hawking--Hartle basis are finally found accordingly as a function of the optical scalars in the Kinnersley tetrad and of the scalar function $\mathcal{A}$ (see for example \cite{Chandrasekhar:1985kt}, p.55; see also \cite{Nerozzi:2016kky} for an illuminating discussion on the topic of spin coefficients and gauge fixing).

The main features of the solutions of the evolution equations for the expansion and the twist, \eqref{eq18} and \eqref{eq19}, are well captured in perturbation theory. To proceed in this direction, a comment is due. The advantage of using the Hawking--Hartle basis resides in the fact that, on the horizon and in the stationary~(not perturbed) configuration of the Kerr black hole, both $\rho$ (and henceforth $\theta$ and $\omega$) and $\sigma$ vanish, while $\epsilon=\epsilon_0$ is constant. However, that is not the case for the whole set of optical scalars: in particular, $\alpha, \beta, \pi$ and $\tau$ are neither zero nor constant on the horizon (they explicitly depend on the polar angle, but they are constant in the time parameter $v$). While it is clearly possible to try to pursue a perturbative solution of \eqref{eq18} and \eqref{eq19} in a completely general fashion, to present a crystal-clear calculation we will make a further simplifying assumption and consider $\kappa_{_T}$ and $\delta^* \kappa_{_T}$ to contribute only at the second order in the perturbative expansion. In this way, it follows that also $A_{_T}$ and $B_{_T}$ enter only at the second perturbative order. Following the arguments of \cite{Chandrasekhar:1985kt}, we arrive at (apices in parentheses denote hereafter the perturbation order) 
\begin{eqnarray}\label{eq:22}
    D \theta^{(1)} &=& 2 \theta^{(1)} \epsilon_0 \,,\nonumber\\
        D \omega^{(1)} &=& 2 \omega^{(1)} \epsilon_0 \,;
\end{eqnarray}
in a rotating black hole, $\theta^{(1)}$ and $\omega^{(1)}$ must be periodic along the horizon generators \cite{Hawking:1972hy}; therefore, the only possible solution of  \eqref{eq:22} is $\theta^{(1)} =\omega^{(1)}= 0$. 
Perturbing \eqref{eq18} at the second order, then, yields
\begin{eqnarray} \label{eq:25}
    D \theta^{(2)} = - |\sigma^{(1)}|^2 + A_{_T}^{(2)} + 2 \epsilon_0 \theta^{(2)} \ ,
\end{eqnarray}
whose solution can be written as
\begin{eqnarray} \label{eq:39}
    \theta^{(2)} = - \int_v^{\infty} e^{2 \epsilon_0 (v-v')} \left( - |\sigma^{(1)}(v')|^2 + A_{_T}^{(2)}(v') \right) dv' \ .
\end{eqnarray}
In the final state, when perturbation relaxes, $\theta^{(2)}$ is expected to vanish; this intuitive claim is corroborated by the form of the solution above. 
The quantity $\theta^{(2)}$ is additionally a measure of convergence of the null congruence emanating from the surface element of the horizon $d \Sigma$,
\begin{eqnarray}
   \theta^{(2)} = \frac{1}{2 d \Sigma} \frac{d (d\Sigma)}{dv}\,.
\end{eqnarray}
For a perturbation starting at an initial time $v_{_i}$, that can be chosen to be equal to zero, and lasting till some time $v_{_f}$, the previous expression can be written in integral form as
\begin{eqnarray}
    \int^{v_{_f}}_{0} \frac{d (d\Sigma)}{d \Sigma} = 2 \int^{v_{_f}}_{0} \theta^{(2)} (v) dv \ .
\end{eqnarray}
Plugging the expression~\eqref{eq:39} in the equation above, we get
\begin{eqnarray}\label{eq:44}
    \log \Bigg[\frac{d \Sigma \vert_{v_{_f}}}{d \Sigma \vert_{0}} \Bigg] &=& - 2 \int^{v_{_f}}_{0} dv  \int_v^{\infty} dv' e^{2 \epsilon_0 (v-v')} \left( - |\sigma^{(1)}(v')|^2 + A_{_T}^{(2)}(v') \right) \nonumber\\
   &=& \frac{1}{\epsilon_0} \int^{v_{_f}}_{0} (1- e^{- 2 \epsilon_0 v} ) \left( |\sigma^{(1)}(v')|^2 - A_{_T}^{(2)}(v') \right) dv \nonumber  \\
   && + \frac{1}{\epsilon_0} (e^{2 \epsilon_0 v_{_f}} - 1) 
    \int^{\infty}_{v_{_f}}  \left( |\sigma^{(1)}(v')|^2 - A_{_T}^{(2)}(v') \right) e^{- 2 \epsilon_0 v}  dv \,. 
\end{eqnarray}
The second term on the right-hand side of the above expression is teleological in nature as the change in horizon area depends on what happens after $v_{_f}$ when the perturbation is already relaxed. Thus, we consider
\begin{eqnarray}
    |\sigma^{(1)}(v)|^2=0=A_{_T}^{(2)} \ , \quad \textrm{for} \quad v > v_{_f} \quad \textrm{and} \quad v_{_f} \gg \frac{1}{2 \epsilon_0} \ .
\end{eqnarray}
Under these assumptions, Eq.~\eqref{eq:44} becomes
\begin{eqnarray}
    \log \Bigg[\frac{d \Sigma \vert_{v_{_f}}}{d \Sigma \vert_{0}} \Bigg] \simeq \frac{1}{\epsilon_0} \int^{v_{_f}}_{0} \left( 
    |\sigma^{(1)}(v')|^2 - A_{_T}^{(2)}(v') \right) dv \ .
\end{eqnarray}
Assuming finally that the change in $d \Sigma$ during the interval $(0, v_{_f})$ is very small yields
\begin{eqnarray}
    \delta (d \Sigma) = \frac{d \Sigma}{\epsilon_0} \int^{v_{_f}}_{0} \left( |\sigma^{(1)}(v')|^2 - A_{_T}^{(2)}(v') \right) dv \ ,
\end{eqnarray}
or, in differential form, 
\begin{eqnarray}\label{boh}
    \frac{d \left(\delta (d \Sigma)\right)}{dv} = \frac{d \Sigma}{\epsilon_0} \left( |\sigma^{(1)}(v')|^2 - A_{_T}^{(2)}(v') \right) \ .
\end{eqnarray}
For a Kerr black hole, the horizon area element is $d \Sigma = 2  M r_+ d \Omega$, where $r_+$ is the radius of the horizon, $\textrm{M}$ is the mass of the black hole, and $d \Omega= \sin \theta d \theta d \phi$ is the solid angle element. In this case, \eqref{boh} reads
\begin{eqnarray} \label{eq32}
    \frac{d \left(\delta (d \Sigma)\right)}{dv d \Omega} = \frac{2 \text{M} r_+}{\epsilon_0} \left( |\sigma^{(1)}(v')|^2 -A_{_T}^{(2)}(v') \right) \ .
\end{eqnarray}
The expression above suggests that the net effect of a non-zero torsion is to contribute to the rate of change in the horizon area already associated with a non-zero shear. However, the sign of the total contribution of the corrections with origin in the torsion dynamics, $A_{_T}^{(2)}$, will be in general dependent on the underlying gravitational theory and on the gravity-matter coupling.

To close this section, it is instructive to solve the equation for the twist at the second order. From 
\begin{eqnarray}
    D \omega^{(2)} = 2 \omega^{(2)} \epsilon_0 + B_{_T}^{(2)} \ ,
\end{eqnarray}
one gets the solution
\begin{eqnarray} \label{eq:35}
    \omega^{(2)} = - \int_v^{\infty} e^{2 \epsilon_0 (v-v')} B_{_T}^{(2)}(v') dv' \ .
\end{eqnarray}
The expression above clearly shows that torsion (or better, the spin-2 component encoded in the $\kappa_{_T}$ term in  $B_{_T}^{(2)}$) generates twist. The twist goes to zero when torsional perturbations die off. As a side comment, it is rather interesting to note that, if the simplifying assumption on $\kappa_{_T}$ and $\delta^* \kappa_{_T}$ is dropped, contorsion could in principle source twist already at the first perturbative order, through $B_{_T}^{(1)}$. This case would be even more dramatic, since a nonvanishing $(\omega^{(1)})^2$ term would now enter in \eqref{eq:25}, providing a second torsion-dependent contribution to the evolution of the expansion. This also proves our calculation to be consistent with what has been predicted in~\cite{Dey:2017fld} considering the approach {\em à la} Jacobson.

\section{Discussion and conclusions\label{sec5}}

The tidal heating phenomenon is a well-known mechanism induced by the deformation of the generators of a black hole horizon under the influence of metric perturbations. The typical astrophysical setting where tidal heating is at work is the in-spiral phase of the evolution of a binary system, when the tidal field of a compact object deforms its companion, and the mass and angular momentum of both members of the system evolve with time. 

In this paper, we have extended the Hawking--Hartle tidal heating formula to a $U_4$ spacetime. 

We have followed the Newman--Penrose tetrad formalism and used optical scalars equations to describe the evolution of a congruence of autoparallel trajectories in the presence of torsion. It was first shown in~\cite{Dey:2017fld} that, when torsion is switched on, the hypersurface orthogonal congruence does not guarantee the twist to vanish. It was conjectured therein that the change in the horizon area due to a torsion-induced perturbation does depend not only on the generated shear but also on the twist. In this article, we have derived the rate of change of the horizon area and shown explicitly that an extra term arises due to torsion. 

In Einstein--Cartan (with a non-minimal coupling between torsion and matter fields, intended to turn on the spin-2 component of the former), we can easily imagine a relevant case where our scheme applies: a high-density neutron star plunging radially into a black hole. The neutron-rich matter in the star, having intrinsic spin, generates torsion and excites the horizon during the merging process.  This excitation is expected to go to zero as the flow of matter-carrying spin extinguishes. From the macroscopic point of view, this could lead to a different merger-related gravitational waves signal associated with the accretion of the neutron star onto a black hole. This in turn might lead to an observably different Bondi mass propagated at infinity.

Furthermore, according to Onsager's principle and following the arguments suggested by Candelas and Sciama, as recalled in our introduction, one can also conclude that the modifications to the Hawking–Hartle formula induced by the presence of torsion will affect (net of the effect of induced four-fermion interactions, see \cite{Flachi:2015fna,Flachi:2014jra,Flachi:2015sva}) the corresponding Hawking radiation. Explicit quantitative predictions in this regard will be pursued elsewhere. Note finally that the idea suggested in~\cite{PhysRevLett.38.1372} is the result of a straightforward application of Onsager's principle to black hole thermodynamics. A covariant generalisation of Onsager's principle, in the context of a geometric formulation of statistical mechanics, see for example \cite{Chirco:2014naa,Chirco:2015bps,Chirco:2017wgl}, while beyond the scope of our program, would surely be of benefit to the present discussion.

\section{Acknowledgement}
VV warmly thanks Andrea Nerozzi for a colourful discussion. SH acknowledges the institutional support of the Research Centre for Theoretical Physics and Astrophysics, Institute of Physics, Silesian University in Opava.
SL acknowledges funding from the Italian Ministry of Education and Scientific Research (MIUR) under the grant PRIN MIUR 2017-MB8AEZ.
The work of VV has been carried out in the framework of activities of the Italian National Group of Mathematical Physics (GNFM, INdAM) and the INFN Research Project QGSKY.

\appendix

\section*{Appendix: Basic definitions  \label{app_bas}}
The covariant derivative for a generic four-vector~$X^{\beta}$ is defined as
\begin{equation} \label{eq:1}
\widehat{\nabla}_{\alpha} X^{\beta} = \partial_{\alpha} X^{\beta} + \Gamma_{\alpha \sigma}{}^{\beta} X^{\sigma} \ .
\end{equation}
In our considerations, the connection $\Gamma_{\alpha \sigma}^{\,\,\,\,\,\,\beta}$ has the only constraint of being metric compatible i.e. 
\begin{equation} \label{eq:2}
\widehat{\nabla}_{\alpha} g_{_{\beta \gamma}} = 0  \quad (\textrm{zero non-metricity}) \ .
\end{equation}
The torsion tensor is defined as the anti-symmetric part of the generic connection
\begin{equation} \label{eq:3}
{T_{\alpha \beta}}^{\gamma} \equiv \Gamma_{[\alpha \beta]}{}^{\gamma} = \frac{1}{2} \left( \Gamma_{\alpha \beta}{}^{\gamma} - \Gamma_{ \beta\alpha}{}^{\gamma} \right) \ .
\end{equation}
The general metric compatible connection can be written as the difference between the Christoffel symbol of the 2nd kind $\genfrac\{\}{0pt}{}{\gamma}{\alpha\beta}$ and the contorsion tensor ${K_{\alpha \beta}}^{\gamma}$,
\begin{equation} \label{eq:5}
\Gamma_{\alpha \beta}{}^{\gamma} = \genfrac\{\}{0pt}{}{\gamma}{\alpha\beta} - {K_{\alpha \beta}}^{\gamma} \ ,
\end{equation}
(following \cite{Jogia1980, Jogia81}, we adopt the minus sign convention in front of ${K_{\alpha \beta}}^{\gamma}$) where
\begin{equation} \label{eq:6}
{K_{\alpha \beta}}^\gamma \equiv -{T_{\alpha \beta}}^\gamma + T_{\beta \,\,\,\alpha}^{\,\,\,\gamma} - {T^\gamma}_{\alpha \beta}  \,,
\end{equation}
is anti-symmetric with respect to the last two indices.

The Ricci identity, finally, reads
\begin{equation} \label{RicciId}
(\widehat{\nabla}_{\alpha} \widehat{\nabla}_{\beta}-\widehat{\nabla}_{\beta}\widehat{\nabla}_{\alpha}) A^\gamma =  - \widehat{R}_{\alpha\beta\lambda}{}^\gamma A^\lambda+ (K_{\alpha \beta}{}^\lambda-K_{\beta\alpha }{}^\lambda)\widehat{\nabla}_{\lambda} A^\gamma   \,,
\end{equation}
where the Riemann curvature of the full connection is defined by
\begin{equation}
    \widehat{R}_{\alpha\beta\lambda}{}^\gamma=-\partial_\alpha \Gamma_{\beta\lambda}{}^\gamma+\partial_\beta \Gamma_{\alpha\lambda}{}^\gamma-\Gamma_{\alpha\mu}{}^\gamma\Gamma_{\beta\lambda}{}^\mu+\Gamma_{\beta\mu}{}^\gamma\Gamma_{\alpha\lambda}{}^\mu\,,
\end{equation}
which is an antisymmetric tensor with respect to the first two and the last two indices.
\bibliography{main}

\begin{thebibliography}{10}

\bibitem{PhysRev.37.405}
L~Onsager.
\newblock {\em Phys. Rev.}, 37:405--426, 1931.

\bibitem{PhysRev.38.2265}
L~Onsager.
\newblock {\em Phys. Rev.}, 38:2265--2279, 1931.

\bibitem{PhysRev.83.34}
H~B Callen and T~A Welton.
\newblock {\em Phys. Rev.}, 83:34--40, 1951.

\bibitem{Kubo:1957mj}
R~Kubo.
\newblock {\em J. Phys. Soc. Jap.}, 12:570--586, 1957.

\bibitem{PhysRevLett.38.1372}
P~Candelas and D~W Sciama.
\newblock {\em Phys. Rev. Lett.}, 38:1372--1375, 1977.

\bibitem{Hartle:1973zz}
J~B Hartle.
\newblock {\em Phys. Rev. D}, 8:1010--1024, 1973.

\bibitem{Hawking:1972hy}
S~W Hawking and J~B Hartle.
\newblock {\em Commun. Math. Phys.}, 27:283--290, 1972.

\bibitem{Hughes:2001jr}
S~A Hughes.
\newblock {\em Phys. Rev. D}, 64:064004, 2001.
\newblock [Erratum: Phys.Rev.D 88, 109902 (2013)].

\bibitem{Alvi:2001mx}
K~Alvi.
\newblock {\em Phys. Rev. D}, 64:104020, 2001.

\bibitem{Chatziioannou:2012gq}
K~Chatziioannou, E~Poisson, and N~Yunes.
\newblock {\em Phys. Rev. D}, 87(4):044022, 2013.

\bibitem{Isoyama:2017tbp}
S~Isoyama and H~Nakano.
\newblock {\em Class. Quant. Grav.}, 35(2):024001, 2018.

\bibitem{Datta:2019epe}
S~Datta, R~Brito, S~Bose, P~Pani, and S~A Hughes.
\newblock {\em Phys. Rev. D}, 101(4):044004, 2020.

\bibitem{Datta:2019euh}
S~Datta and S~Bose.
\newblock {\em Phys. Rev. D}, 99(8):084001, 2019.

\bibitem{Maselli:2017cmm}
A~Maselli, P~Pani, V~Cardoso, T~Abdelsalhin, L~Gualtieri, and V~Ferrari.
\newblock {\em Phys. Rev. Lett.}, 120(8):081101, 2018.

\bibitem{Datta:2020gem}
S~Datta, K~S Phukon, and S~Bose.
\newblock {\em Phys. Rev. D}, 104(8):084006, 2021.

\bibitem{Datta:2020rvo}
S~Datta.
\newblock {\em Phys. Rev. D}, 102(6):064040, 2020.

\bibitem{Datta:2021row}
S~Datta and K~S Phukon.
\newblock {\em Phys. Rev. D}, 104(12):124062, 2021.

\bibitem{Chakraborty:2021gdf}
S~Chakraborty, S~Datta, and S~Sau.
\newblock {\em Phys. Rev. D}, 104(10):104001, 2021.

\bibitem{Chandrasekhar:1985kt}
S~Chandrasekhar.
\newblock {\em {The mathematical theory of black holes}}.
\newblock Oxford University Press, 1983.

\bibitem{Dey:2017fld}
R~Dey, S~Liberati, and D~Pranzetti.
\newblock {\em Phys. Rev. D}, 96(12):124032, 2017.

\bibitem{Dey:2022fll}
S~Dey and B~R Majhi.
\newblock arXiv:2206.11875 [gr-qc], 2022.

\bibitem{Dey:2022qqp}
S~Dey and B~R Majhi.
\newblock {\em Phys. Rev. D}, 105(6):064047, 2022.

\bibitem{Jacobson:1995ab}
T~Jacobson.
\newblock {\em Phys. Rev. Lett.}, 75:1260--1263, 1995.

\bibitem{Luz:2017ldh}
P~Luz and V~Vitagliano.
\newblock {\em Phys. Rev. D}, 96(2):024021, 2017.

\bibitem{Iosifidis:2018diy}
D~Iosifidis, C~G Tsagas, and A~C Petkou.
\newblock {\em Phys. Rev. D}, 98(10):104037, 2018.

\bibitem{Cembranos:2018ipn}
J~A~R Cembranos, J~Gigante~Valcarcel, and F~J Maldonado~Torralba.
\newblock {\em JCAP}, 04:039, 2019.

\bibitem{Hensh:2021oyk}
S~Hensh and S~Liberati.
\newblock {\em Phys. Rev. D}, 104(8):084073, 2021.

\bibitem{Jogia1980}
S~{Jogia} and J~B {Griffiths}.
\newblock {\em General Relativity and Gravitation}, 12(8):597--617, 1980.

\bibitem{Jogia81}
S~Jogia.
\newblock {\em A spin-coefficient approach to space–times with torsion}.
\newblock PhD thesis, Loughborough University, 1981.

\bibitem{Speziale:2018cvy}
S~Speziale.
\newblock {\em Phys. Rev. D}, 98(8):084029, 2018.

\bibitem{Luz:2019kmm}
P~Luz and F~C Mena.
\newblock {\em J. Math. Phys.}, 61(1):012502, 2020.

\bibitem{test1}
S~Speziale.
\newblock Private communication.

\bibitem{Hehl:1976kj}
F~W Hehl, P~Von Der~Heyde, G~D Kerlick, and J~M Nester.
\newblock {\em Rev. Mod. Phys.}, 48:393--416, 1976.

\bibitem{Magueijo:2012ug}
J~Magueijo, T~G Zlosnik, and T~W~B Kibble.
\newblock {\em Phys. Rev. D}, 87(6):063504, 2013.

\bibitem{Benedetti:2011nd}
D~Benedetti and S~Speziale.
\newblock {\em JHEP}, 06:107, 2011.

\bibitem{Vitagliano:2010pq}
V~Vitagliano, T~P Sotiriou, and S~Liberati.
\newblock {\em Phys. Rev. D}, 82:084007, 2010.

\bibitem{Vitagliano:2010sr}
V~Vitagliano, T~P Sotiriou, and S~Liberati.
\newblock {\em Annals Phys.}, 326:1259--1273, 2011.
\newblock [Erratum: Annals Phys. 329, 186--187 (2013)].

\bibitem{Vitagliano:2013rna}
V~Vitagliano.
\newblock {\em Class. Quant. Grav.}, 31(4):045006, 2014.

\bibitem{Kinnersley:1969zza}
W~Kinnersley.
\newblock {\em J. Math. Phys.}, 10:1195--1203, 1969.

\bibitem{Nerozzi:2016kky}
A~Nerozzi.
\newblock {\em Phys. Rev. D}, 95(6):064012, 2017.

\bibitem{Flachi:2015fna}
A~Flachi.
\newblock {\em Int. J. Mod. Phys. D}, 24:1542017, 2015.

\bibitem{Flachi:2014jra}
A~Flachi and K~Fukushima.
\newblock {\em Phys. Rev. Lett.}, 113(9):091102, 2014.

\bibitem{Flachi:2015sva}
A~Flachi, K~Fukushima, and V~Vitagliano.
\newblock {\em Phys. Rev. Lett.}, 114(18):181601, 2015.

\bibitem{Chirco:2014naa}
G~Chirco, C~Rovelli, and P~Ruggiero.
\newblock {\em Class. Quant. Grav.}, 32(3):035011, 2015.

\bibitem{Chirco:2015bps}
G~Chirco, T~Josset, and C~Rovelli.
\newblock {\em Class. Quant. Grav.}, 33(4):045005, 2016.

\bibitem{Chirco:2017wgl}
G~Chirco, D~Oriti, and M~Zhang.
\newblock {\em Phys. Rev. D}, 97(12):126002, 2018.

\end{thebibliography}
\bibliographystyle{unsrt}
\end{document}